\documentclass[prc,article,twocolumn,showpacs,amssymb]{revtex4}

\usepackage{graphics}
\usepackage{epsfig}

\begin{document}

\title{Behavior of odd-even mass staggering around $^{132}$Sn}

\author{L. Coraggio$^{1}$, A. Covello$^{1,2}$, A. Gargano$^{1}$,
and N. Itaco$^{1,2}$} 
\affiliation{$^{1}$Istituto Nazionale di Fisica Nucleare, 
Complesso Universitario di Monte S. Angelo, I-80126 Napoli,
Italy\\
$^{2}$Dipartimento di Fisica, Universit\`a
di Napoli Federico II,
Complesso Universitario di Monte S. Angelo,  I-80126 Napoli,
Italy}

\date{\today}

\begin{abstract}

We have performed shell-model calculations of binding energies of nuclei around 
$^{132}$Sn.  The main aim of our study has been to find out if the behavior of odd-even staggering across $N=82$ is explainable in terms of the shell model.
In our calculations, we have employed realistic low-momentum two-body
effective interactions derived from the charge-dependent Bonn nucleon-nucleon
potential that have already proved quite successful in describing the 
spectroscopic properties of nuclei in the $^{132}$Sn region. Comparison shows 
that our results fully explain the trend of the experimental staggering.

\end{abstract}    

\maketitle

A fundamental property of an atomic nucleus is its mass. Mass spectrometry studies started more than a century ago and since then have continued achieving higher and higher accuracy.
A comprehensive historical overview of mass spectrometry since the very beginning is given in Ref. \cite{Blaum06}. 

Today, the highest accuracy (parts per billion) is reached with ion traps,
in particular Penning traps. With this technique, mass measurements on neutron-rich Sn and Xe isotopes were recently performed \cite{Dworschak08,Neidherr09}  at CERN On-Line Isotope Mass Separator (ISOLDE) facility by using the mass spectrometer ISOLTRAP. A remarkable result of the study \cite{Dworschak08} was that the mass measurement of $^{134}$Sn revealed a 0.5-MeV discrepancy with respect to previous $Q_\beta$ measurements. This provided clear evidence of the robustness of the $N=82$ shell closure, ruling out the hypothesis of an $N=82$ shell quenching.
 
The $^{132}$Sn region is currently the  focus of great experimental and theoretical
interest, especially in view of the production of new neutron-rich nuclear species 
at the next generation of radioactive ion beam facilities. 
Very recently, atomic masses of several neutron-rich nuclei around $^{132}$Sn have been measured  \cite{Hakala12} using the JYFLTRAP Penning trap mass spectrometer coupled to the Ion Guide Isotope Separator On-Line (IGISOL) facility at the accelerator laboratory of the University of
Jyv{\"a}skyl{\"a} . 
In this study, the masses of some nuclei, as for instance $^{135}$Sn and $^{136}$Sb, were measured for the first time and the precision of previously measured masses was significantly improved. Attention was also focused on the odd-even staggering (OES) of binding energies
for $N=81$ and 83 isotones. In particular, the experimental values for Sn, Te, and Xe   were compared with those obtained by performing various state-of-the-art self-consistent calculations with the SLy4 Skyrme energy density functional and contact pairing force. 
No calculation of Ref.~\cite{Hakala12}, however, was able to  reproduce the experimental behavior of the staggering  in the $N=83$ isotones. This led the authors to consider this behavior anomalous and attribute it to specific effects 
beyond the $N=82$ shell gap  not accounted for, in their opinion,  by current theoretical approaches.

Over the past several years we have conducted several shell-model studies  of neutron-rich nuclei around $^{132}$Sn
\cite{Coraggio09a,Covello11,Danchev11,Coraggio13a} by using  Hamiltonians with
single-particle and single-hole energies taken from experiment and effective
interactions derived from the charge dependent (CD-Bonn) nucleon-nucleon ($NN$) potential
renormalized through the $V_{low-k}$ procedure \cite{Bogner02} with a cutoff momentum $\Lambda$ of
2.2 fm$^{-1}$. All these studies, focused essentially on the energy spectra and
electromagnetic properties, led to results in very good agreement with
experiment. 
The findings mentioned above have challenged us to put our realistic shell-model calculations to the test also in this puzzling case.

In our calculations, we assume that the valence protons and the valence neutron holes occupy the five orbits $0g_{7/2}$, $1d_{5/2}$, $1d_{3/2}$, $0h_{11/2}$, and $2s_{1/2}$ of the 50-82 shell
while the valence neutrons have available the six orbits $1f_{7/2}$, $2p_{3/2}$, $2p_{1/2}$, 
$0h_{9/2}$, $1f_{5/2}$, and  $0i_{13/2}$ of the 82-126 shell. The adopted values of the single-particle and single-hole energies are reported in Refs. \cite{Coraggio13a} and
\cite{Danchev11}, respectively. They were taken from the experimental spectra of 
$^{133}$Sb \cite{ENSDF}, $^{131}$Sn \cite{ENSDF,Fogelberg04}, and $^{133}$Sn \cite{ENSDF} with the exceptions of the proton $2s_{1/2}$ and the neutron $0i_{13/2}$ energies which were from Refs. \cite{Andreozzi97} and \cite{Coraggio02}, respectively,  since the corresponding single-particle levels are still missing in the spectra of $^{133}$Sb and $^{133}$Sn. We should also point out here that, as in our most recent calculations \cite{Covello11,Coraggio13a}, the experimental energy of  \cite{Jones10} is used for the neutron  $2p_{1/2}$ level. The needed mass excesses are taken from Ref. \cite{Hakala12}.

As mentioned above, the two-body effective interaction $V_{\rm eff}$ is derived from the CD-Bonn $NN$ potential, whose  short-range repulsion is renormalized by means of the
$V_{\rm low-k}$ potential \cite{Bogner02} with $\Lambda=2.2$ fm$^{-1}$. The 
obtained low-momentum potential is then used, with the addition of the Coulomb force for protons, to derive $V_{\rm eff}$ within the framework of a perturbative approach based on the $\hat Q$-box folded-diagram expansion \cite{Coraggio09b,Coraggio12}. Some details on the calculation of the two-body interaction above and below $N=82$ can be found in \cite{Coraggio13b} and \cite{Danchev11}, respectively.

\begin{table}
\caption{ Calculated and experimental binding energies $B$, relative to $^{132}$Sn, of Sn, Sb, Te, and Xe isotopes. } 

\begin{ruledtabular}
\begin{tabular}{ccc}
Nucleus & $B_{\rm calc}$ & $B_{\rm expt}$\\
& (MeV)& (MeV)\\
\colrule
$^{134}$Sn & 5.98 & 6.03\\
$^{135}$Sn & 8.37 & 8.30\\
$^{134}$Sb & 12.74 & 12.84\\
$^{135}$Sb & 16.32 & 16.58\\
$^{136}$Sb & 18.81 & 19.47\\
$^{134}$Te & 20.81 & 20.57\\
$^{135}$Te & 23.82 & 23.83\\
$^{136}$Te & 28.26 & 28.60\\
$^{137}$Te & 31.15 & 31.55\\
$^{136}$Xe & 40.32 & 39.04\\
$^{137}$Xe & 43.83 & 43.06\\
$^{138}$Xe & 48.89 & 48.73\\
\end{tabular}
\end{ruledtabular}
\label{tabsp}
\end{table}

To start with, we report in Table I the calculated binding energies, relative to $^{132}$Sn, of the Sn, Sb, Te, and Xe isotopes beyond $N=82$ and compare them with 
the results from the mass measurements performed in \cite{Hakala12} for the first three kinds of isotopes and in 
\cite{Neidherr09} for the latter. Note that the errors  on the measured values are in the order of keV and therefore are not given in Table I, where the reported energies are rounded to tens of keV.
We may also  mention that the values of the binding energies reported in our previous papers   (see, for instance, Refs. \cite{Covello11,Coraggio05,Coraggio06}) for some of  these nuclei differ slightly  from the present ones. This is because here  we use  a different energy value  for the the neutron  $2p_{1/2}$ level and the mass excesses measured in  \cite{Hakala12}. From Table I,  we see that the experimental data are remarkably well reproduced by the theory, the largest discrepancy not exceeding 3\%. It should be emphasized
that for all 12 nuclei considered we have used a unique Hamiltonian with a realistic two-body effective interaction containing no free parameters.

Making use of the binding energies of $^{134}$Sn, $^{134-136}$Te and $^{136-138}$Xe
reported in Table I and of those obtained for $^{130}$Sn, $^{132-133}$Te and $^{134-135}$Xe,
we have calculated the neutron OES, as given by the three-point formula 
\cite{Satula98,Bertsch09}

\begin{eqnarray}
\Delta^{(3)}(N,Z) = \frac{1}{2} [B(N+1,Z) + B(N-1,Z)  -2B(N,Z)],    \nonumber
\end{eqnarray}

\noindent for the $N=81$ isotones $^{131}$Sn, $^{133}$Te, and $^{135}$Xe and for the $N=83$  
isotones $^{133}$Sn, $^{135}$Te, and $^{137}$Xe. 

\begin{figure}
\vspace{0.25cm}
\begin{center}
\includegraphics [scale=0.52,angle=0] {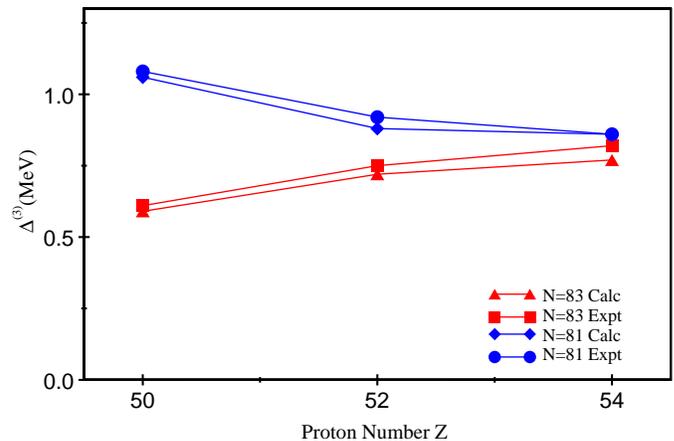}
\end{center}
\caption{\label{fig1} (Color online) Calculated and experimental odd-even staggering for the $N=81$ and 83 isotones. }
\end{figure}

In Fig. \ref{fig1} we compare the calculated OES values  with the experimental ones. We see that the agreement between theory and experiment is very good.
In particular, our calculations quantitatively describe  the gap  between the $N=81$ and 83  lines at $Z=50$  as well as  its  decrease when adding two and four protons, 
which confirms the reliability of the various components of our effective interactions.

The drop of about 0.5 MeV in the observed OES for Sn when crossing $N=82$  is accounted for by the different pairing properties of our effective interaction for
neutron particles and holes with  respect to the $N=82$ closed shell. In fact, the $J^{\pi}=0^{+}$ matrix elements, which are the only ones  entering  the calculation of the ground-state energies of  $^{134}$Sn and $^{130}$Sn, are overall less attractive for the former.
For instance,   the $J^{\pi}=0^+$ diagonal matrix element for the $(1f_{7/2})^{2}$ configuration, which dominates the ground-state wave function of $^{134}$Sn,  is  $-0.65$ MeV, namely about 0.5 MeV less attractive than that for the $(0h_{11/2})^{-2}$ configuration,  whose role is very relevant to the ground state of $^{130}$Sn. In previous works \cite{Coraggio09a,Covello13} we have investigated  the microscopic origin of the paring force above the $N=82$ shell  within our  derivation of the effective interaction. We have analyzed the various perturbative contributions and found that the reduction of the pairing component
 is due to the minor role played by the one particle-one hole excitations, which are instead responsible for a ``normal" pairing below this shell.
It is worth mentioning that the difference in  the pairing force across $N=82$ was also shown to be crucial in reproducing the asymmetric behavior of the yrast $2^+$ state  in tin and tellurium isotopes with respect to  $N=82$ \cite{Terasaki02,Shimizu04}.

 When going to Te and Xe, the $N=81$ and 83  lines
come  closer to each other  as a result of the proton-neutron effective interaction. The two lines  would be indeed  parallel should one ignore this interaction.
From Fig. \ref{fig1}, we see that the $p-n$ interaction has an opposite effect on the $N=81$ and $N=83$ isotones, which is clearly related to its repulsive and attractive nature 
 in the particle-hole and particle-particle channel, respectively. On the other hand,
this effect is  not very large  either in  $^{133,135}$Te or in  $^{135,137}$Xe, since it  results essentially from the difference between the contributions of the $p-n$ interaction to the energies of the  odd  and neighboring  even isotopes. It  makes, however,  the OES almost equal 
 in $^{135}$Xe and $^{137}$Xe, as  is experimentally observed.
 
In summary, we have shown that there is no anomaly in the OES of binding energies
around $^{132}$Sn, as it is fully explained in terms of the shell model with realistic effective interactions which are just the same as those  employed in our previous studies \cite{Coraggio09a,Covello11,Danchev11} in the $^{132}$Sn region. 
\begin{acknowledgments}
This work has been supported in part by the Italian Ministero dell'Istruzione, dell'Universit\`{a} e  della Ricerca (MIUR) under PRIN 2009.
\end{acknowledgments}

\end{document}